\documentclass[]{article}
\usepackage{emulateapj}
\usepackage{psfig}
\begin{document}

\title{General Relativistic Constraints on Emission Models of\\
  Anomalous X-ray Pulsars}

\author{Simon DeDeo, Dimitrios Psaltis, and Ramesh Narayan}

\affil{Harvard-Smithsonian Center for Astrophysics, 60 Garden
St., Cambridge, MA 02138; sdedeo, dpsaltis, narayan@cfa.harvard.edu}


\begin{abstract}
  Most models of anomalous X-ray pulsars (AXPs) account for the
  observed X-ray spectra and pulsations by means of radiation
  processes that occur on the surfaces of neutron stars. For any such
  model, general relativistic deflection of light severely suppresses
  the amplitude of the observed pulsations. We calculate the expected
  pulsation amplitudes of AXPs according to various models and compare
  the results with observations. We show that the high ($\lesssim
  70$\%) pulse amplitudes observed in some AXPs can be accounted for
  only if the surface emission is localized (spot radius $<40^\circ$)
  and strongly beamed ($\sim \cos^n\theta'$ with $n\gtrsim 2$, where
  $\theta'$ is the angle to the normal). These constraints are
  incompatible with those cooling and magnetar models in which the
  observed X-rays originate as thermal emission from the neutron-star
  surface. Accretion models, on the other hand, are compatible with
  observations for a wide range of parameters. Finally, definitive
  conclusions cannot be reached on magnetospheric models, since their
  localization and beaming properties are not well understood.
\end{abstract}


\centerline{Submitted to {\em The Astrophysical Journal\/}.}

\section{INTRODUCTION}

Among pulsating compact X-ray sources are a small subset with
pulsation periods between 6 and 12 seconds, soft spectra, and no
identifiable companions. The first suggestion that these objects might
form a separate class of neutron stars -- later called the anomalous
X-ray pulsars (AXPs) -- was made by Mereghetti \& Stella (1995), who
proposed that they might be powered by accretion from a very low mass
companion (this was also earlier hinted at by Hellier 1994). The lack
of optical counterparts, however, and the absence of observable
Doppler shifts in the frequency of the X-ray pulses led van Paradijs,
Taam, \& van den Heuvel (1995) to favor a different accretion model
for AXPs.  According to their suggestion, material from a fossil
accretion disk, possibly the debris of a disrupted binary companion
after a period of common-envelope evolution, is being accreted by a
solitary neutron star. Recently, a similar model has been proposed by
Chatterjee et al.\ (2000), in which the accreting material is supplied
by the post-supernova fallback material from the neutron star
progenitor itself.

In a different class of models, AXPs are considered to be isolated
neutron stars, spinning down by magnetic dipole radiation. Because of
their unusually high period derivatives, a simple application of the
relationship between the dipole magnetic field strength and the
spin-down rate implies a field strength for these objects of $\sim
10^{14}-10^{15}$~G (see, e.g., Thompson \& Duncan 1996). Two main
types of models that rely upon the presumed high magnetic field of the
stars have been proposed.  Thompson \& Duncan (1996; see also Duncan
\& Thompson 1992) suggested that the released energy may be drawn from
the decay of the magnetic field itself and from differential movements
in the stellar crust. This model also serves to explain the bursts
observed from soft gamma-ray repeaters as being produced by
larger-scale magnetospheric phenomena. Because of such models, AXPs
are often called ``magnetars'' and are grouped into the same class of
sources as the soft gamma-ray repeaters (see, e.g, Thompson \& Duncan
1995, 1996; also Hurley 2000 for a review of SGR properties and
models). In the alternative high magnetic-field model, Heyl \&
Hernquist (1997a, 1997b) suggested that AXPs draw their energy from
the residual thermal energy of the star itself.

All these models face a number of difficulties. For example, if AXPs
are powered by accretion from a stellar companion, the absence of
detectable Doppler shifts in the arrival of X-ray pulses cannot be
easily explained(Mereghetti et al.\ 1998). If accretion is from either
a companion or a fossil disk, the optical fluxes one would expect
directly from the disk or due to X-ray reprocessing are too high
compared to the observed upper limits (see, e.g., Perna, Hernquist, \&
Narayan 2000; Hulleman et al.\ 2000). On the other hand, in the
magnetar model the absence of bright faster AXPs and the observed
variations in spin-down rates are hard to account for (see, e.g.,
Baykal \& Swank 1996; Marsden et al.\ 2000; see, however, Heyl \&
Hernquist 1999; Melatos 1999).

In all the above models of AXP, nearly all of the X-ray emission is
produced at the surface of the neutron star. It is well known that
strong gravitational fields tend to smooth out the variability
produced by a spinning compact star, even if the emission is highly
localized in bright spots (Pechenick, Ftaclas, \& Cohen 1983). Indeed,
the X-ray pulse amplitudes of the three radio pulsars that show
thermal emission from their surfaces are only $\lesssim 30$\% (see,
e.g., the discussion in Page 1995; Harding \& Muslimov 1998). This is
in contrast to the non-thermal emission from radio pulsars (which is
magnetospheric in origin) and from accretion-powered X-ray pulsars
(which is from collimated accretion columns) which often show pulse
amplitudes as high as $\sim 90$\% (see, e.g., Nagase 1989). In this
respect, AXPs are similar to accretion-powered pulsars, showing X-ray
pulse amplitudes anywhere between $\sim 10$\% and $\sim 70$\%.

We study in this paper a set of variability diagnostics that may be
used in constraining emission models of AXPs. We examine three
parameters: the pulse fraction observed at infinity, which is a
measure of the overall amplitude of the variations, and the Fourier
amplitudes at the first and second harmonics of the neutron-star spin
frequency.  We find tight constraints on the properties of magnetar
models. Most importantly, we are unable to reproduce the observed
variability properties of AXPs with thermal cooling models.

\section{FORMALISM}

In order to determine how relativistic effects suppress or enhance
variability amplitudes, we need to consider curved photon paths from
the surface of the star to an observer at infinity. Since the objects
under consideration are rotating slowly, we use the Schwarzschild
spacetime, which is appropriate for a non-spinning mass, and ignore
effects such as relativistic frame dragging, which are important only
for rapidly spinning objects. In this section, we outline the basic
ingredients of our method, drawing on the work of Pechenick et al.\
(1983).

For each model we specify both the brightness distribution over the
surface of the star and the effective beaming of radiation, i.e., we
specify the specific intensity integrated over photon energy,
$I(\theta,\phi,\theta')$, that emerges from each point on the stellar
surface with polar coordinates $(\theta,\phi$) at an angle $\theta'$
with respect to the normal.

In the current analysis we consider a number of different, physically
motivated, mathematical expressions for the dependence of the emerging
specific intensity on $\theta'$ (hereafter called the beaming
function). This allows us to explore a large parameter space and draw
conclusions that do not depend strongly on any particular emission
model. We consider isotropic emission, i.e., no dependence on
$\theta'$, as well as the beaming described by the Hopf function
(Chandrasekhar 1950, eq.~[III.50])
\begin{equation}
I(\theta,\phi,\theta')=I_0(\theta,\phi)
   \left(\sum_{a=1}^3 \frac{L_a}{1+k_a\cos{\theta'}}-\cos\theta'+Q\right)\;,
\end{equation}
where the parameters $L_a$, $k_a$, and $Q$ are given in Chandrasekhar
(1950; Table~III.VII). The Hopf function describes the beaming of
radiation emerging from a scattering atmosphere heated from below and
is, therefore, suitable for a weakly-magnetic H--He atmosphere at
energies $\gtrsim 1$~keV (Zavlin et al.\ 1998). 

For photon energies near the cyclotron energy ($E_{\rm cyc}\simeq
11.6[B/10^{12}$~G$]$~keV), the beaming function for a magnetic
atmosphere does not always decrease monotonically away from the radial
direction. Indeed the beaming function may have a local minimum at
small angles from the normal (Zavlin et al.\ 1995).  However, for a
dipole magnetic field this effect is not very significant and results
in a rather small pulse fraction ($\lesssim 30$\%) even when general
relativistic effects are not taken into account (see Zavlin et al.\
1995). Moreover, the cooling models of AXPs discussed here 

\vbox{ \centerline{ \psfig{file=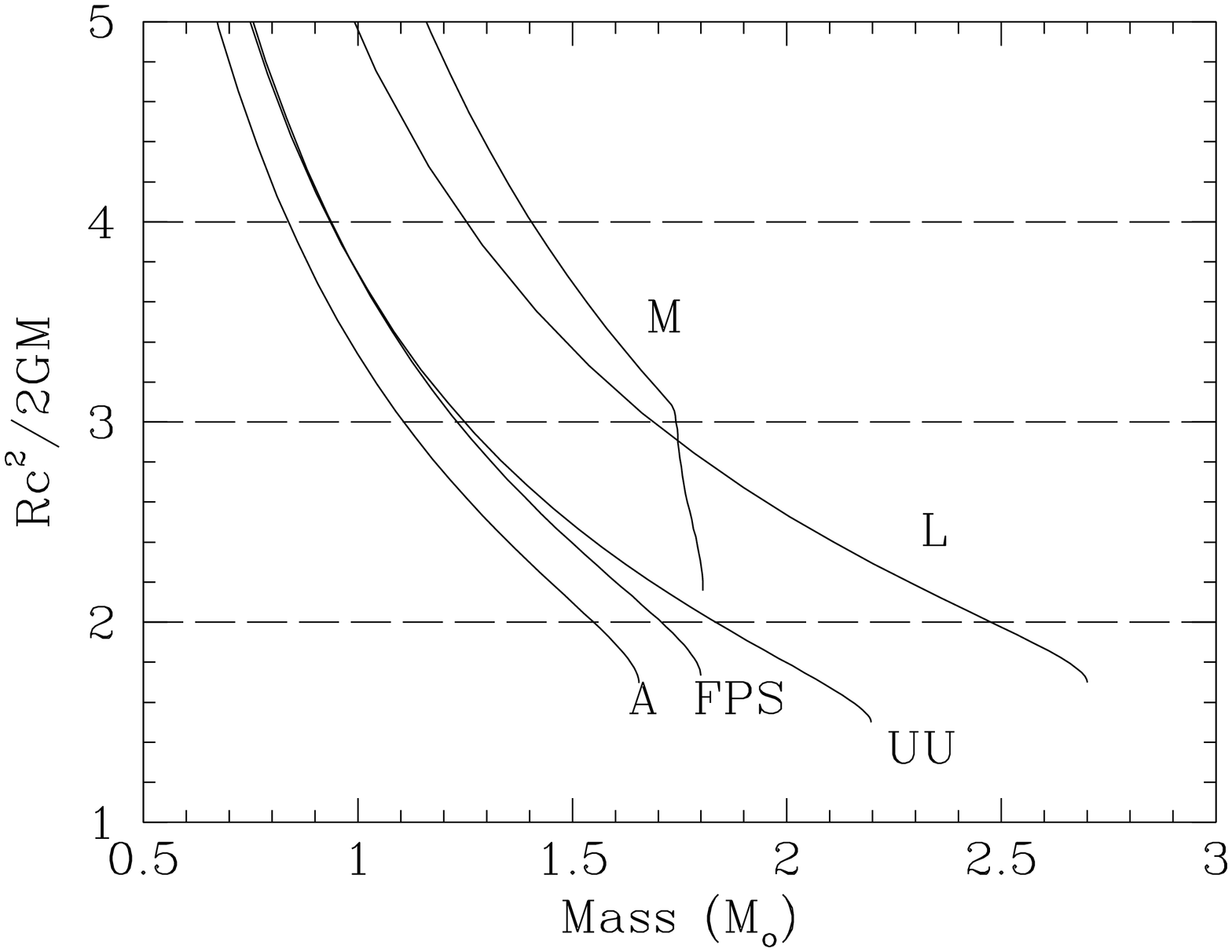,angle=-90,width=9.0truecm} }
\figcaption[]{\footnotesize Neutron star radii, in units of the
    Schwarzchild radius, for different neutron star masses and
    equations of state (A: Pandharipande 1971; FPS: Lorenz, Ravenhall,
    \& Pethick 1983; UU: Wiringa, Fiks, \& Frabrocini 1988; L:
    Pandharipande \& Smith 1975b; M: Pandharipande \& Smith 1975a).}}
\vspace*{0.5cm}

\noindent require that the magnetic field of the neutron star is
$\sim 10^{14}-10^{15}$~G and hence that the cyclotron energy is $\sim
1-10$~MeV, i.e, much larger than the $\sim 1$~keV photon energies that
are of interest here. For such energies, it is reasonable to consider
beaming functions that decrease monotonically away from the normal to
the surface (Zavlin et al.\ 1995).

In models of accretion columns, the interaction of radiation with a
plasma in the strong magnetic field of the neutron star, as well as
the possibility of radiation being obscured by the accretion column
(see, e.g., Riffert et al.\ 1993) leads to a much sharper beaming.
For this reason, we also consider beaming functions of the form
\begin{equation}
I(\theta,\phi,\theta')=I_0(\theta,\phi)\cos^n\theta'\;.
\label{eq:cosn}
\end{equation}
According to Nagel (1981; see also Meszaros \& Nagel 1985), the
radiation pattern emerging from an accretion column at low accretion
rates is described quite well with $n\simeq 2-3$.

Given a model for the X-ray emission from the stellar surface, the
flux measured by an observer at distance $d$, whose polar coordinates
with respect to the stellar rotation axis are $(\beta,\Phi)$, is given
by (cf.\ Pechenick et al.\ 1983)
\begin{eqnarray}
  F_{\infty}(\beta,\Phi)&= &\left(\frac{R}{d}\right)^2
    	\left(\frac{M}{R}\right)^2 
	\left(1-\frac{2M}{R}\right)^2\nonumber\\
   & & \times \int_{x=0}^{x_{\rm
    	max}}\int_{y=0}^{2\pi}I(\theta,\phi,\theta')\, x\, dx\, dy\;.
\label{eq:Finfty}
\end{eqnarray}
In the above equation $\theta$, $\phi$, and $\theta'$ depend
implicitly on the angles $x$ and $y$ as described in Pechenick et al.\
(1983), $x_{\rm max}\equiv(R/M)(1-2M/R)^{-1/2}$, $M$ and $R$ are the
neutron star mass and radius, and we have set $c=G=1$. The double
integral has an integrable pole at $x_{\rm max}$.  In the calculations
presented here we have evaluated this integral to an accuracy of
$10^{-3}$ using Romberg integration of the fifth order.

\noindent\vbox{
\begin{tabular}{clcc}
\multicolumn{4}{c}{AXP PROPERTIES}\\
\hline
Label & Source Name & Pulse Fraction & $I_2/I_1$\\ 
\hline
A & 1E~1048.1--5937 & 0.76 & 0.15\\ 
B & 1E~1841--045 & 0.15 & 0.54\\ 
C & AX~J1845.0--0258 & 0.63 & 0.14\\
D & 1RXS J170849.0$-$400910 & 0.50 & 0.40\\ 
E & 4U~0142$+$61 & 0.17 & 0.70\\ 
F & 1E~2259$+$586 & 0.35 & 1.35\\
\hline
\label{computed} 
\end{tabular}}
{\footnotesize References.---\ A: Oosterbroek et al.\ 1998 ({\em
SAX\/}); B: Gotthelf et al.\ 1999 ({\em ASCA+SAX\/}); C: Torri et al.\
1998 ({\em ASCA\/}); D: Sugizaki et al.\ 1997 ({\em ASCA\/}); E:
Israel et al.\ 1999 ({\em SAX\/}); F: Iwasawa et al.\ 1992 ({\em
GINGA}).}

\vspace{0.5cm}

The degree of suppression of the pulsation amplitude depends
sensitively on the compactness of the neutron star. Figure~1 shows the
ratio $p\equiv Rc^2/2GM$ for different neutron-star masses and
equations of state. Based on this figure, we limit our parameter study
to $p=2,3$, and 4; larger values of $p$ would correspond to
unrealistically light neutron stars, even for the stiffest proposed
equations of state (i.e., $\lesssim 1.4 M_\odot$ even for equation of
state M).

\section{PULSATION AMPLITUDES}

The pulsation amplitudes observed in anomalous X-ray pulsars allow us
to place constraints on models of X-ray emission from their
surface. In this paper, we do not attempt to fit particular observed
pulse profiles but rather try to set general constraints on 
large classes of models. For this reason, we only consider the pulse
fraction, defined as (cf.\ eq.[\ref{eq:Finfty}])
\begin{equation}
PF\equiv \frac{F_\infty^{\rm max}-F_\infty^{\rm min}}
   {F_\infty^{\rm max}+F_\infty^{\rm min}}\;,
   \label{eq:PF}
\end{equation}
and the Fourier amplitudes of the harmonics of the pulse frequency,
defined by
\begin{equation}
F_\infty(\beta,\Phi)= I_0(\beta)+I_1(\beta) \cos(\Phi)+
   I_2(\beta)\cos(2\Phi) +...
   \label{eq:harm}
\end{equation}
Table~I shows the observed pulse fractions and harmonic content for
six AXPs, as inferred approximately from published lightcurves.  We
see that half of the known systems have pulse fractions of 0.5 or
larger, with the largest being $\sim 0.7$. As we discuss below, this
fact provides strong constraints on some models of AXPs.

\subsection{Cooling of Magnetic Neutron Stars}

The brightness distribution on the surface of a strongly-magnetic
cooling neutron star depends on the local magnetic field strength $B$
and its angle $\psi$ with respect to the local radial direction (see,
e.g., Heyl \& Hernquist 1998). The flux emerging from a spot on the
stellar surface is $\sim B^m \cos^2\psi$, with $m\simeq 0.4$
approximating well the numerical results (Heyl \& Hernquist 1998). For
a dipole stellar field we, therefore, use
\begin{equation}
I_0(\theta,\phi)\sim I_0 \frac{\cos^2\theta}{(3\cos^2\theta+1)^{1-m/2}}\;,
   \qquad m=0.4,
\label{eq:HH}
\end{equation}
where ($\theta$,$\phi$) are polar coordinates on the stellar surface
with respect to the magnetic axis. 

\vbox{ \centerline{ \psfig{file=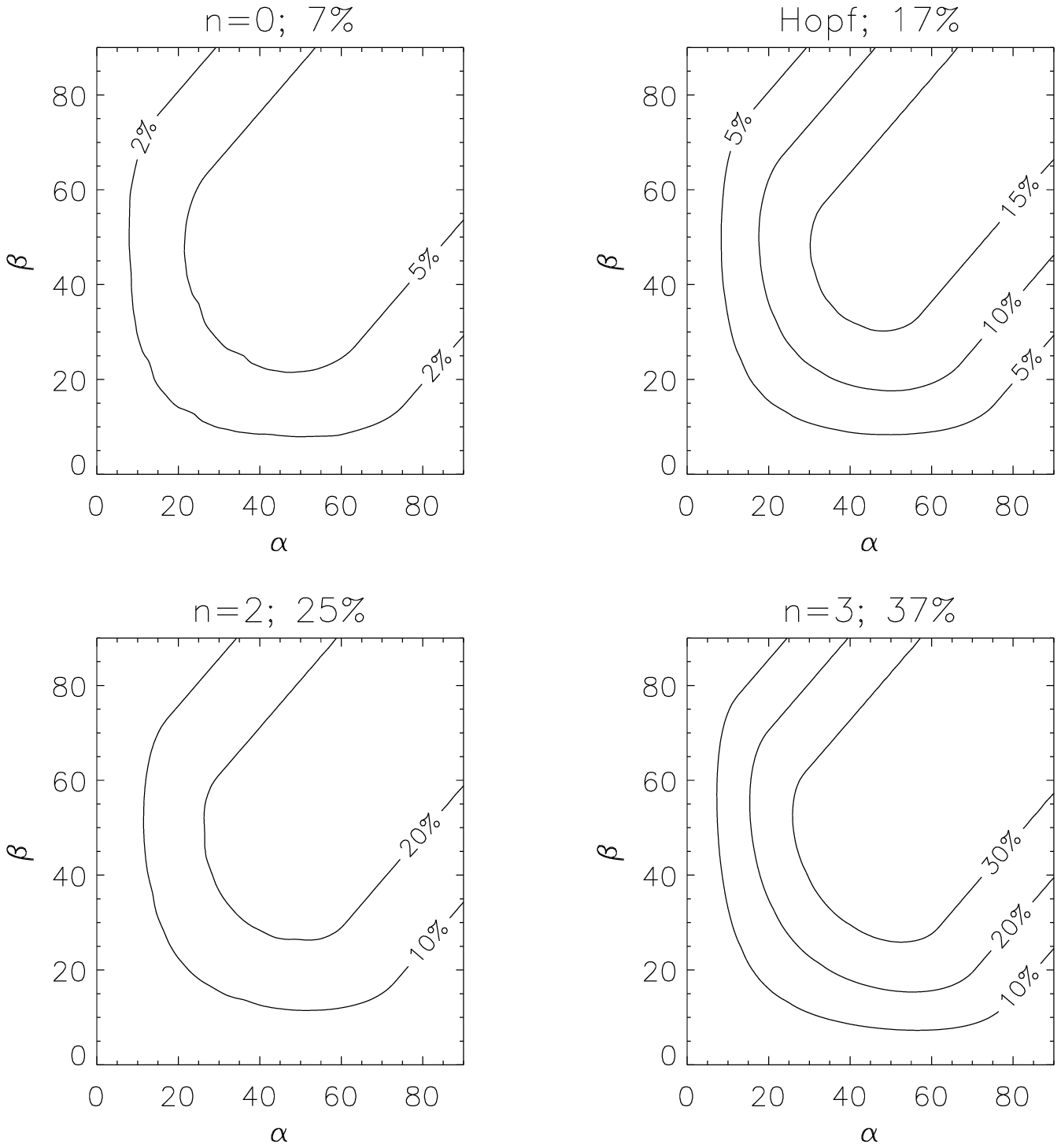,angle=0,width=9.0truecm} }
\figcaption[]{\footnotesize Pulse fractions calculated for cooling
   neutron stars with a dipole magnetic field, and different magnetic
   inclinations ($\alpha$) and orientations of the observer ($\beta$)
   with respect to the rotation axis. The different panels correspond
   to an isotropic beaming function ($n=0$), a Hopf function, and
   beaming functions that are increasingly more peaked towards the
   radial direction ($n=2,3$).}}
\vspace{0.5cm}

Figure~2 shows the predicted pulse fractions for a cooling
neutron-star model, for different choices of the angle $\alpha$
between the magnetic dipole axis and the rotation axis and the angle
$\beta$ between the light of sight to the observer and the rotation
axis, as well as for different beaming functions. Some general
features of this figure are worth noting. For example, the predicted
pulse fraction remains unchanged when the magnetic inclination and the
inclination of the observer are interchanged. This is true because the
flux at infinity measured at any give pulse phase $\Phi$ depends only
on the angular distance
\begin{equation}
   \theta_0=\cos^{-1}(\sin\alpha\sin\beta\cos\Phi+\cos\alpha\cos\beta)
\end{equation}
between the magnetic axis and the direction to the observer, which is
symmetric in $\alpha$ and $\beta$. Furthermore, the predicted pulse
fraction is maximum when $\alpha=\beta$. The overall maximum occurs
when $\alpha=\beta=90^\circ$ and the pulse fraction is zero when
either $\alpha=0^\circ$ or $\beta=0^\circ$.

The most important result from Figure~2 is that even for the most
favorable geometry (i.e., $\alpha=\beta=90^\circ$) and for strong
beaming (i.e., $n=3$), the pulse fraction does not exceed 37\%. The
reason is that the surface brightness distribution~(\ref{eq:HH}) is
too smooth.  Assuming a less relativistic neutron star ($R/2M=4$), or
even a significantly stronger dependence of the emerging flux on the
local magnetic field ($m=1$ in eq.~[\ref{eq:HH}]) results in only a
modest increase of the maximum pulse fraction (51\% and 50\%,
respectively, for $n=3$). We thus conclude that neutron star cooling
models with the surface emission described by equation~(\ref{eq:HH})
cannot reproduce the large pulse fractions observed from AXPs.

\vbox{ \centerline{ \psfig{file=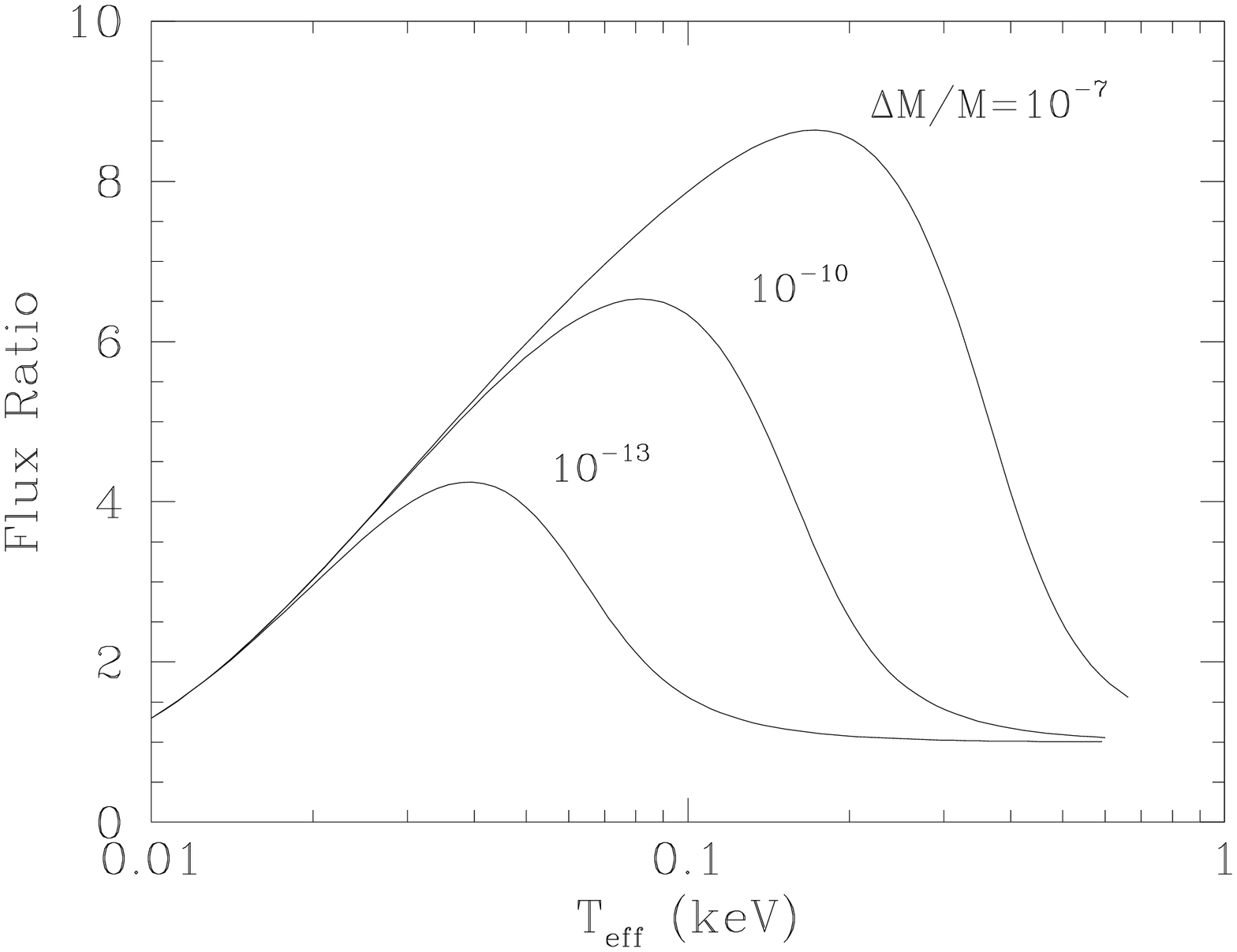,angle=-90,width=9.0truecm} }
\figcaption[]{\footnotesize Ratio of the bolometric flux emerging
   from an atmosphere with a given mass fraction $(\Delta
   M/M)(\Omega/4\pi)$ (in terms of the neutron-star mass $M$) of
   H--He matter accreted over a fraction $\Omega/4\pi$ of the stellar
   surface, to the bolometric flux from a purely iron atmosphere with
   the same core temperature (after Chabrier et al.\ 1997).}}
\vspace{0.5cm}

\subsection{Localized Thermal Emission}

Thermal emission from the neutron-star surface can, in principle, be
more localized than indicated in equation~(\ref{eq:HH}), if it
is confined mostly around the stellar magnetic poles.  For example,
metallicity gradients on the surface of a cooling neutron star,
possibly produced by magnetically-channelled fallback material during
the supernova explosion, can lead to larger effective temperatures
near the magnetic poles than in the magnetic equator (see, e.g.,
Pavlov et al.\ 2000). Alternatively, non-uniform heating of the
neutron-star atmosphere, e.g., caused by magnetic field decay or
crustal heating (see, e.g., Thompson \& Duncan 1996), can lead to
more localized thermal emission from its surface.

We estimate the magnitude of the first effect using the analytic
expressions for neutron-star atmosphere models given by Chabrier,
Potekhin, \& Yakovlev (1997). We assume that a fraction $(\Delta
M/M)(\Omega/4\pi)$ of light element material has accumulated only over
a fraction ($\Omega/4\pi$) of the stellar surface. We neglect the fact
that the surface layers of the neutron star are in the liquid phase
and hence lateral diffusion may smooth out the metallicity gradient
(L.\ Hernquist, private communication). For a given core temperature
we then calculate the ratio $f$ of the bolometric flux emerging from
the light-element region of the surface to the bolometric flux from
the region consisting purely of iron. This flux contrast is shown in
Figure~3, for different values of the mass fraction of the accreted
atmosphere.  The flux contrast does not further increase, for $(\Delta
M/M)(\Omega/4\pi)\gtrsim 10^{-7}$, because the base of the accreted
layer reaches densities that are high enough ($\sim
10^{10}$~g~cm$^{-3}$) for it to be part of the isothermal
core. Although the above models are strictly valid only for weakly
magnetic neutron stars, the flux ratios are significantly more
sensitive to the composition of the neutron-star envelope than the
magnetic field strength (cf.\ Heyl \& Hernquist 1997b).

\vbox{ \centerline{ \psfig{file=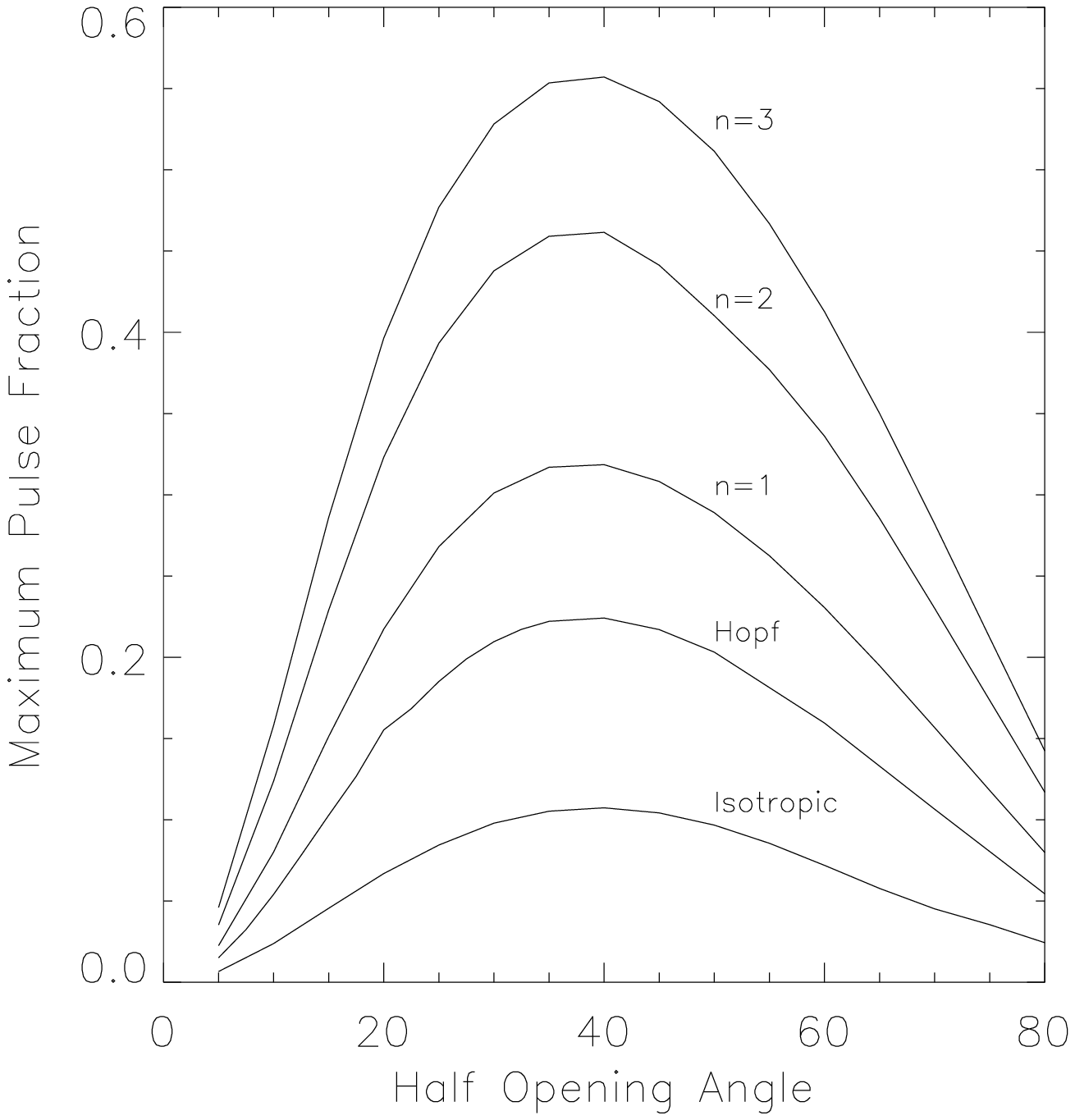,angle=0,width=9.5truecm,height=7.0truecm} }
\figcaption[]{\footnotesize Calculated pulse fractions for a
   neutron star with $R/2M=3$, as a function of the half-opening angle
   of each polar cap, for different beaming functions, and a flux
   contrast of 9 between the polar caps and the rest of the stellar
   surface.}}
\vspace{0.5cm}

According to Figure~3, the expected flux contrast between the polar
caps and the rest of the stellar surface can be at most $f\sim 9$ and
the contrast depends very weakly on the amount of accreted
material. For the flux contrast to attain its maximum value, the polar
caps must accrete $\Delta M\gtrsim 10^{-7} M_\odot$, while the rest of
the surface must accumulate more than three orders of magnitude less
material.  Even if these conditions were met and even if the photons
emerged from the stellar surface strictly radially, the pulse fraction
would be at most $(f-1)/(f+1)\sim 0.8$.

Any realistic beaming function reduces this upper bound significantly
below the highest observed pulse fraction for AXPs ($\sim 0.7$; see
Table~I), even for mildly relativistic neutron stars. This is
demonstrated in Figure~4, which shows the maximum pulse fraction of
radiation emerging from a neutron star with $R/M=3$, as a function of
the angular radius of each polar cap, for various realistic beaming
functions, and an emerging flux from the polar caps that is nine times
larger than the rest of the stellar surface. For small polar-cap
sizes, the radiation flux emerging from the caps is only a small
perturbation to the total brightness of the star and hence the pulse
fraction is small. With increasing cap size the relative contribution
of the caps to the brightness of the star increases, leading to an
increase of the pulse fraction, until the caps cover a large enough
fraction of the stellar surface and the pulse fraction drops
again. For the flux contrast and neutron star radius used here, the
maximum pulse fraction occurs when the polar-cap size is $\sim
40^\circ$ and this maximum value is $\sim 55$\%. Clearly, no realistic
model of this kind can fit the pulse fraction of $\sim 0.7$ seen in
1E~1048.1$-$5937.

The pulse fraction can be further enhanced if the fluxes emerging from
two antipodal polar caps are unequal, e.g., because of uneven fallback
or crustal heating. As the contrast between the two polar caps
increases, the pulse fraction becomes larger, but only at the expense
of the Fourier amplitudes at the harmonics of even order. This is
shown in Figure~5, where the predicted ratio $I_2/I_1$ is plotted
against the pulse fraction for $R/M=3$, two 

\vbox{ \centerline{ 
\psfig{file=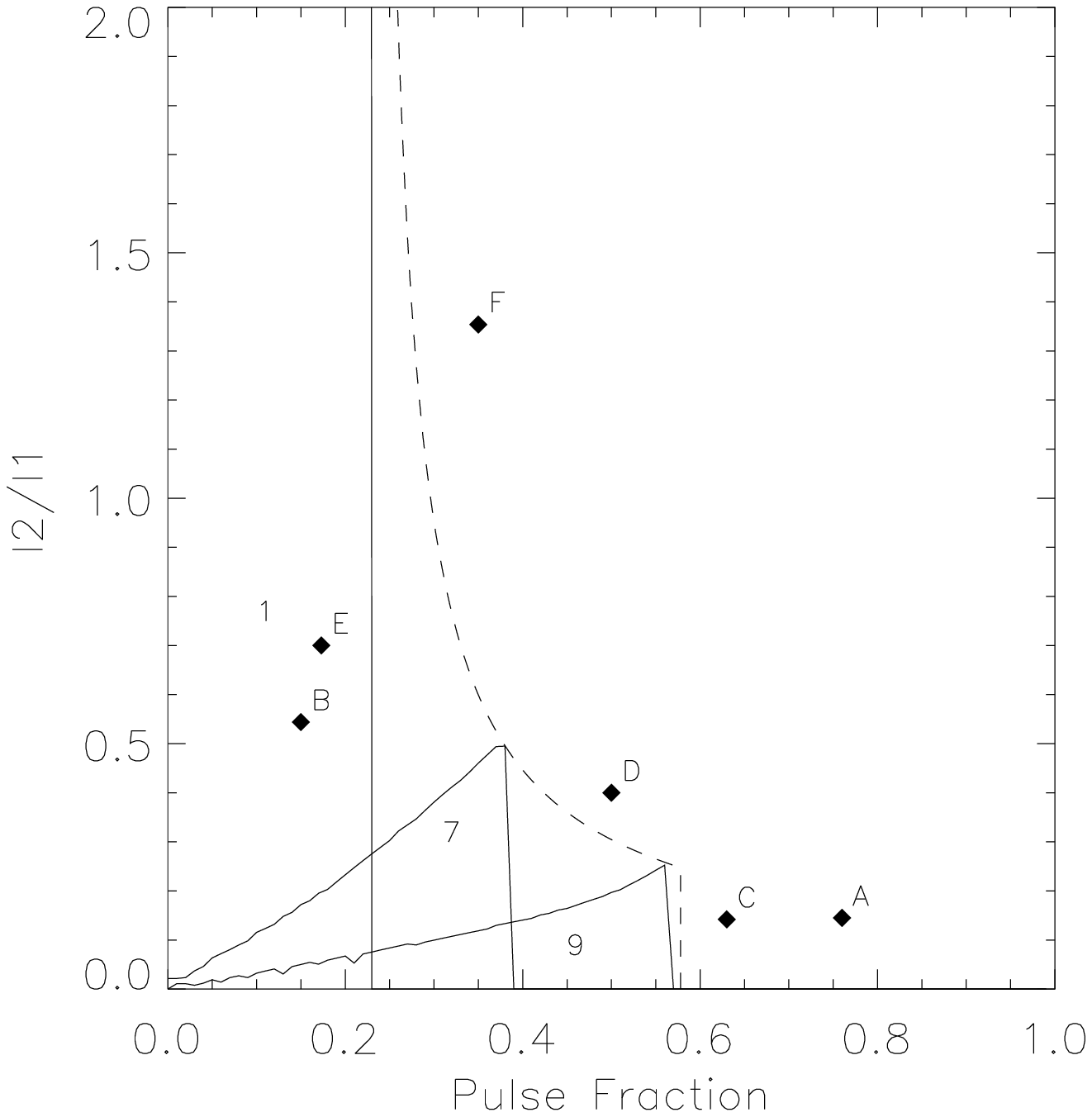,angle=0,width=9.5truecm,height=7.0truecm} }
\figcaption[]{\footnotesize Ratio $I_2/I_1$ of the Fourier
   amplitudes at the second harmonic and fundamental of the pulsar
   spin frequency (cf.\ eq.~[\ref{eq:harm}]) as a function of the
   total pulse fraction (cf.\ eq.~[\ref{eq:PF}]). The calculations
   correspond to $R/M=3$, a configuration of two antipodal polar caps
   with a half-opening angle of 40 degrees but different emerging
   fluxes, and beaming described by the Hopf function. The solid lines
   outline the area in the parameter space allowed for specific ratios
   of fluxes emerging from the two polar caps. The dashed line
   represents the maximum ratio $I_2/I_1$ that corresponds to a given
   pulse fraction; for such a configuration, no system is allowed to
   lie to the right of the dashed line. The diamonds correspond to the
   observed properties of AXPs (cf.\ Table 1).}}
\vspace{0.5cm}

\noindent antipodal caps with a half-opening angle of 40 degrees, and beaming
described by the Hopf function. For these calculations, the two caps
are assumed to have different emerging radiation fluxes, with the flux
of the brightest set to 9 times the flux emerging from the rest of the
neutron-star surface. For any given pulse fraction, there exists a
maximum value that the ratio $I_2/I_1$ can attain, shown by the dashed
line in Figure~5. As a result, detection of a source with a large
pulse fraction and significant amplitude at the even order harmonics
can exclude such a configuration.

The observed properties of some AXPs are not consistent even with a
model with such unequal polar caps (Figure~5). For example, the source
1E~2259$+$586 is characterized by a large pulse fraction and high
harmonic content that cannot be achieved by any of the configurations
considered here.  It does not, therefore, appear plausible for thermal
emission from the stellar surface to account for the variability
properties of AXPs.  Achieving the kind of variability observed is
possible only when the emission is both localized and strongly beamed.

\subsection{Accretion Onto Magnetic Neutron Stars}

In accretion models of AXPs (see, e.g., Mereghetti \& Stella 1995; van
Paradijs et al.\ 1996; Chatterjee et al.\ 1999), a large fraction of
the X-ray emission may be produced mainly at localized ``hot spots''
where the accretion columns meet the stellar surface. In order to take
such configurations into account, we describe the surface brightness
distribution with two circular antipodal spots, with a brightness that
is constant over their surface area.  We denote by $\alpha$ the
angular distance of the center of each spot from its closest rotation
pole.

\vbox{ \centerline{ \psfig{file=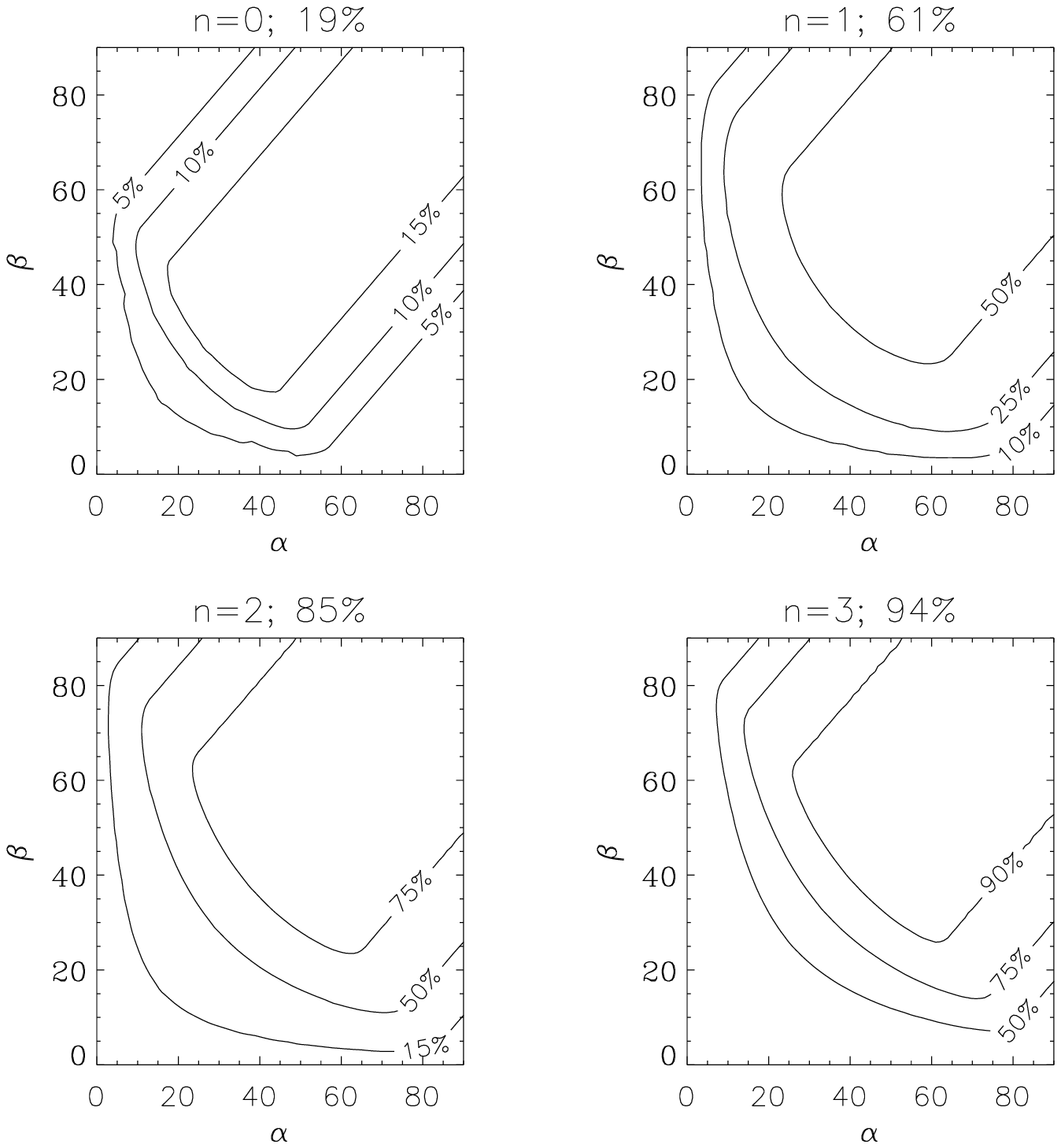,angle=0,width=9.0truecm} }
\figcaption[]{\footnotesize Calculated pulse fractions for a
   neutron star with $R/2M=3$, a hot-spot half-opening angle of 5
   degrees and a $\cos^n\theta'$ beaming function, with n=0--3. The
   maximum value of the pulse fraction is written above each panel.}}

\vspace{0.5cm}

\vbox{ \centerline{ \psfig{file=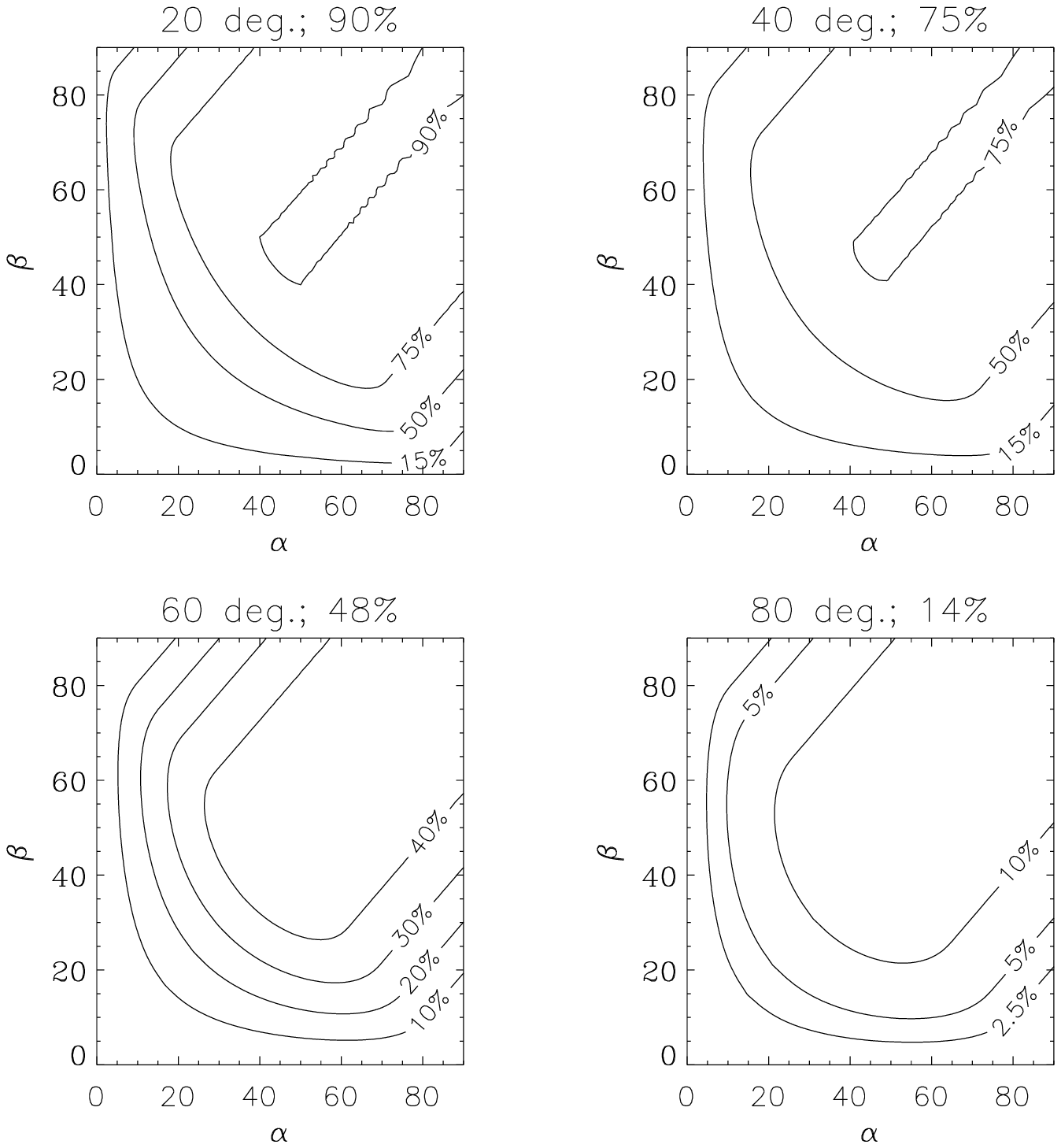,angle=0,width=9.0truecm} }
\figcaption[]{\footnotesize Calculated pulse fractions for a
   neutron star with $R/2M=3$, a $\cos^3\theta'$ beaming function, and
   different hot-spot half-opening angles. The maximum value of the
   pulse fraction is written above each panel.}}

\vspace{0.5cm}

If AXPs are powered by accretion from a geometrically thin disk, the
half-opening angle of each polar cap can be very small. We estimate
the size of each polar cap as the angular distance on the stellar
surface from the magnetic axis of the footpoint of the last magnetic
field line that 

\vbox{ \centerline{ \psfig{file=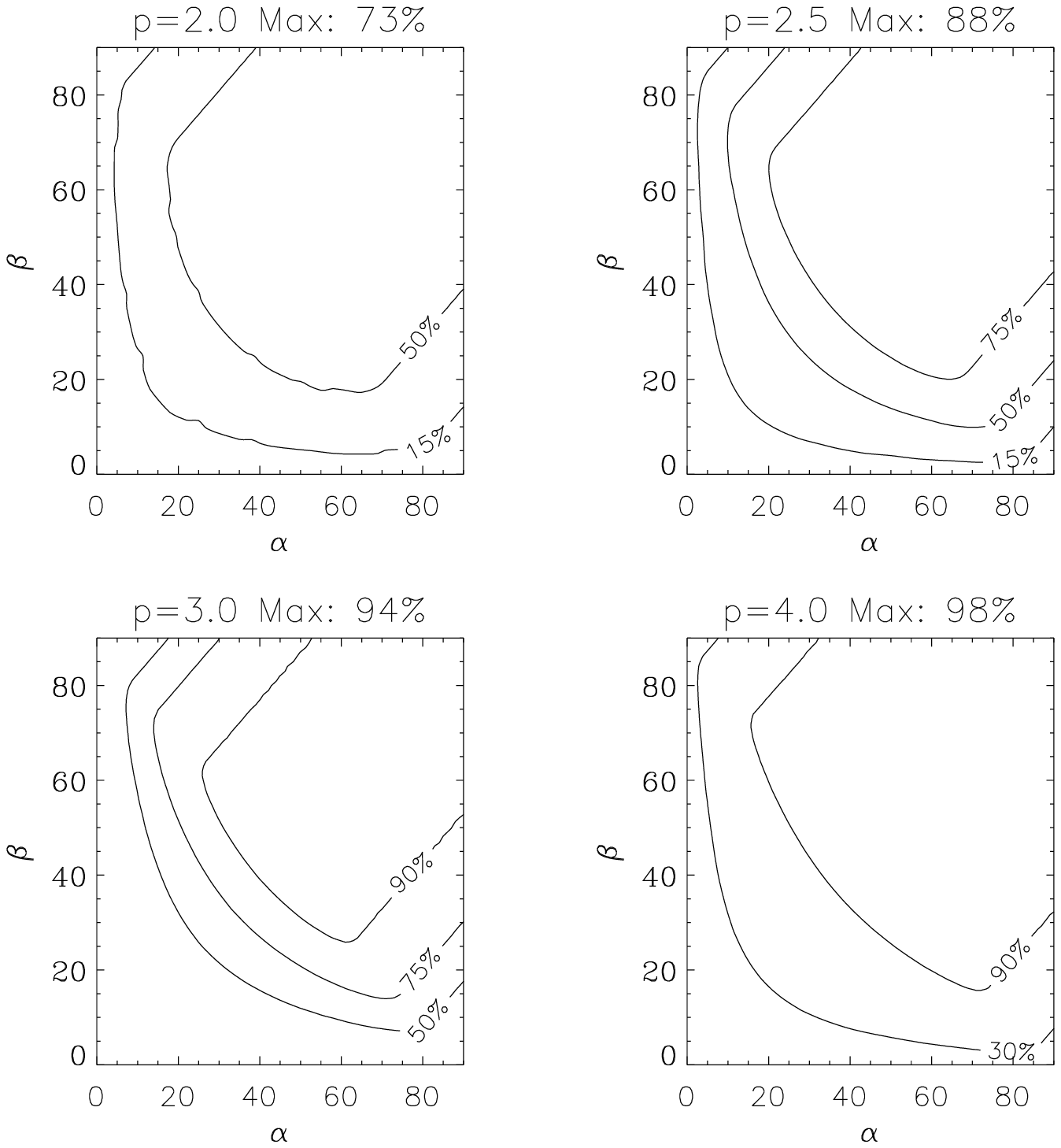,angle=0,width=9.0truecm} }
\figcaption[]{\footnotesize Calculated pulse fractions for
   different neutron-star radii ($p=R/2M$), a hot-spot half-opening
   angle of 5 degrees, and a $\cos^3\theta'$ beaming function. The
   maximum value of the pulse fraction is written above each panel.}}

\vspace{0.5cm}

\noindent penetrates the accretion disk. Given that all AXPs are
observed to be spinning down, their accretion disks must be truncated
near the outer corotation radius (see, e.g., Ghosh \& Lamb 1979)
\begin{eqnarray}
R_{\rm co}&\equiv&\left(\frac{GM P^2}{4\pi^2}\right)^{1/3}\nonumber\\
 &\simeq&
   550\, R\left(\frac{R}{10^6~\mbox{cm}}\right)^{-1}
   \left(\frac{M}{1.4 M_\odot}\right)^{1/3}
   \left(\frac{P}{6~\mbox{s}}\right)^{2/3}\;,
\end{eqnarray}
where $P$ is the spin period of the pulsar. For a dipole magnetic
field, the quantity $\sin^2\theta/r$, with $\theta$ measured from the
magnetic pole, remains constant along a field line, and therefore the
half-opening angle of the polar cap is $\sim (R/R_{\rm co})^{1/2}
\simeq 2.5^\circ$.

Figure~6 shows the pulse fraction calculated for a model with two
identical hot spots, a half-opening angle of 5 degrees, and various
beaming functions (see eq.~[\ref{eq:cosn}]). Because of the general
relativistic deflection of light, we see that even such a small spot
cannot produce a $\sim 70$\% pulse fraction, unless there is
significant beaming ($n\gtrsim 2$). On the other hand, if the emerging
radiation is strongly beamed towards the radial direction, a large
pulse fraction can be achieved for a wide range of polar cap sizes
($\lesssim 40^\circ$; Fig.~7) and for all realistic neutron-star radii
(Fig.~8).

\section{CONCLUSIONS}

In this paper we have used the high pulse amplitudes observed from a
number of AXPs to constrain the properties of their emission
mechanism. We find that the observations can be accounted for only
if the surface emission is localized (half-opening angle $<40^\circ$)
and strongly beamed 

\vbox{ \centerline{ 
\psfig{file=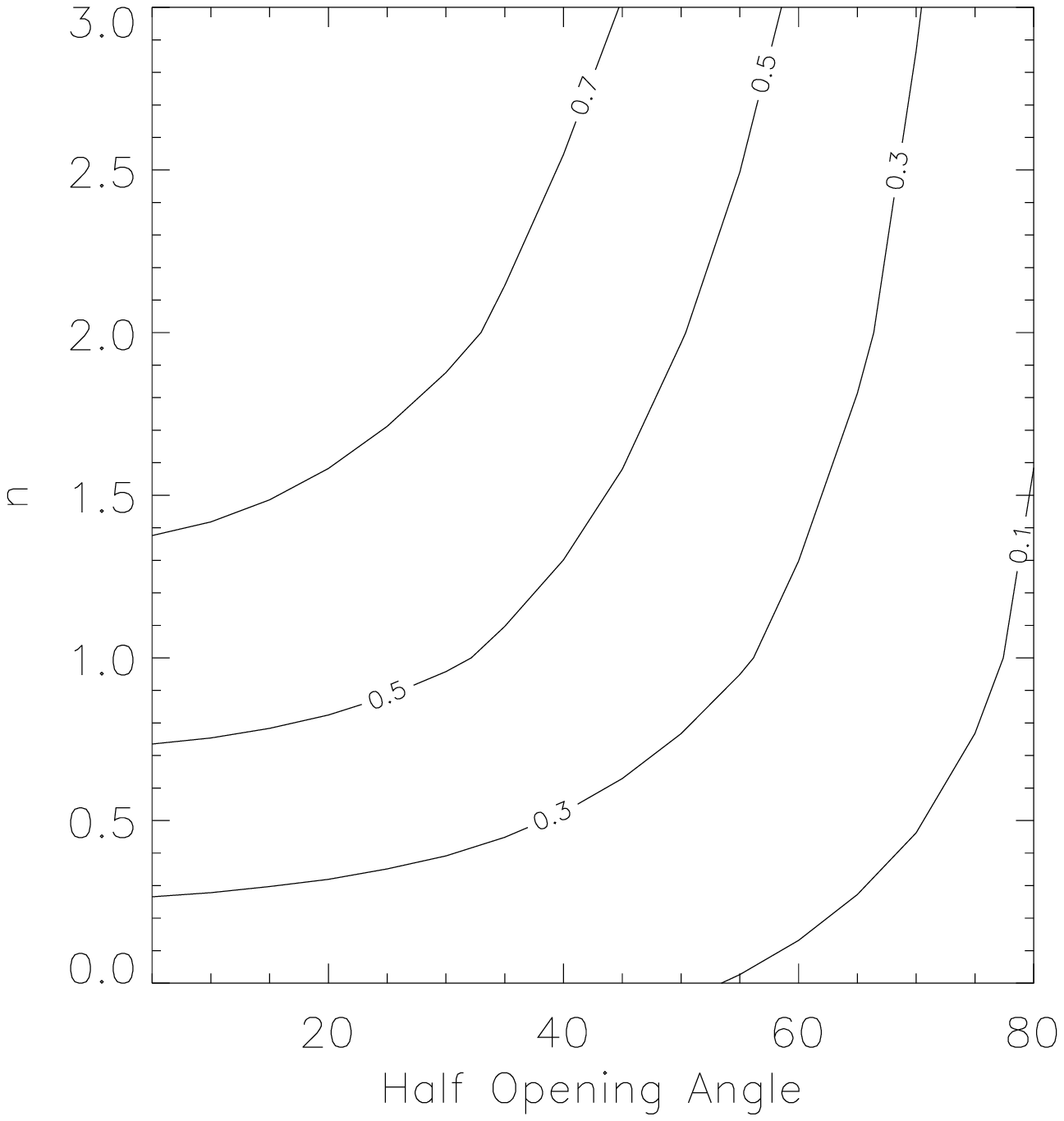,angle=0,width=9.5truecm,height=7.0truecm} }
\figcaption[]{\footnotesize Contours of maximum pulse fraction for
   different beaming functions ($\cos^n\theta'$; cf.\
   eq.[\ref{eq:cosn}]) and half-opening angles of each hot spot. The
   $\simeq 70$\% pulse fraction observed from the source
   1E~1048.1-5937 restricts any viable model of AXPs to the left of
   the uppermost contour.}}
\vspace{0.5cm}

\noindent ($n\gtrsim 2$ in eq.~[\ref{eq:cosn}]), as summarized
quantitatively in Figure~9. These constraints are a consequence of the
compactness of the neutron stars and the resulting strong general
relativistic deflection of photon trajectories. Our conclusions are
valid for all realistic neutron star masses and radii.

The properties of individual sources offer a number of additional
clues. For example, the double-peaked pulse profile of 1E~2259$+$586
requires that the emission is localized around {\em two\/} antipodal
spots on the neutron-star surface, probably associated with the
magnetic poles. Furthermore, the change in the relative strength of
the two peaks observed with {\em GINGA\/} (Iwasawa et al.\ 1992)
implies that the pulse shape cannot be solely due to geometric effects
but should also reflect a flux contrast between the two antipodal
spots. Furthermore, this flux contrast should be variable and,
therefore, cannot be caused by a non-dipolar magnetic field
configuration or a non-uniform fallback of low metallicity material.

Such arguments, together with the constraints presented in Figures~2,
4, 5, and 9 appear to rule out thermal cooling models for AXPs.  They
are also inconsistent with those magnetar models in which most of the
X-ray flux originates from heating in the deep surface layers of the
neutron star. On the other hand, the localized emission and beaming
predicted by accretion models seem to be consistent with the
observations.  A magnetospheric model in which the neutron-star
surface is heated over localized spots by particle bombardment may
also be viable, though the beaming properties of such a model are
unknown.

\acknowledgements

We thank Deepto Chakrabarty, Lars Hernquist, Vicky Kaspi, Jessica
Lackey, and Feryal \"Ozel for many useful discussions. This work was
supported in part by NSF grant AST~9820686. D.\,P.\ acknowledges the
support of a postdoctoral fellowship of the Smithsonian Institution.

\end{document}